\def\version{May 8, 2006; revised July 1, 2006}
\documentclass[aps,prb,twocolumn,superscriptaddress,floatfix]{revtex4}
\usepackage{epsfig,amsmath,amssymb}
\usepackage{hyperref}
\def\abs#1{\vert#1\vert}
\def\EO{EO}
\def\Sztot{S^z}
\begin{document}

\title{Frustrated ferromagnetic spin-$1/2$ chain in a magnetic field: The phase diagram and thermodynamic properties}

\author{F. Heidrich-Meisner}
\affiliation{Materials  Science and Technology Division, Oak Ridge National Laboratory,
 Oak Ridge, Tennessee, 37831, USA and\\
 Department of Physics and Astronomy, University of Tennessee, Knoxville,
 Tennessee 37996, USA}

\author{A. Honecker}
\affiliation{Institut f\"ur Theoretische Physik, Universit\"at G\"ottingen,
%Friedrich-Hund-Platz 1,
37077 G\"ottingen, Germany}
\affiliation{Technische Universit\"at Braunschweig, Institut f\"ur Theoretische Physik,
  38106 Braunschweig, Germany}

\author{T. Vekua}
\affiliation{Universit\'e Louis Pasteur, Laboratoire de Physique Th\'eorique,
% 3 Rue de l'Universit\'e,
67084 Strasbourg Cedex, France}
\affiliation{Andronikashvili Institute of Physics, Tamarashvili 6,
 0177 Tbilisi, Georgia}
 
\date{\version}
                        
%-------------------------------------------------------------------------- 
\begin{abstract}
The frustrated ferromagnetic spin-$1/2$ Heisenberg chain is studied by means of  a low-energy field theory as well as
the density-matrix renormalization group and exact diagonalization methods. Firstly, we study the 
ground-state phase diagram in a magnetic field and find an `even-odd' ({\EO}) phase characterized 
by bound pairs of magnons in the region of two weakly coupled antiferromagnetic chains. 
A jump in the magnetization 
curves signals a first-order transition at the boundary of the {\EO} phase, but otherwise the 
 curves are smooth. Secondly, we discuss thermodynamic properties at zero field, 
where we confirm  a double-peak structure in the specific heat for moderate 
frustrating next-nearest neighbor interactions.
\end{abstract}
%-------------------------------------------------------------------------- 

\maketitle
%************************************************************************** 
% Introduction
%**************************************************************************

The physics of frustrated quantum spin systems is currently attracting  
large interest as exotic quantum phases may emerge.\cite{reviews} Prominent examples
are quantum-disordered ground states with different patterns of broken translational 
symmetry and quantum chiral phases (see, e.g., Ref.~\onlinecite{review-frust-1D}).
In addition, some frustrated systems have a large number of low-lying excitations,  leading 
to unusual features in thermodynamic quantities.

In one dimension, the paradigmatic model
is the frustrated spin-$1/2$ chain:
\begin{equation}
H= \sum_l \lbrack J_1 \vec{s}_l \cdot \vec{s}_{l+1}
 + J_2 \vec{s}_l \cdot \vec{s}_{l+2} \rbrack-h \sum_l s_l^z \, .
\label{eq:1.1}
\end{equation}
$\vec{s}_{l}$ are spin $s=1/2$ operators at site $l$, while $h$
denotes a magnetic field.

Much is known about the ground-state properties and the magnetic phase diagram
of the  frustrated antiferromagnetic (AFM) chain with $J_1,J_2>0$.\cite{review-frust-1D}
We highlight 
the appearance of a plateau in the magnetization curve at
magnetization $M=1/3$ and the existence of an `even-odd' (\EO) region
at small $J_1$ with spins flipping in pairs in a magnetic field
$h$.\cite{okunishi03}

Relatively little attention has been paid to frustrated ferromagnetic (FM)
chains, i.e., $J_1 < 0$ and  $J_2 > 0$, until the recent discovery of
materials described by parameters with this combination of signs. We mention in particular
Rb$_2$Cu$_2$Mo$_3$O$_{12}$ which is believed to be described by
$J_1 \approx -3 \, J_2$,\cite{hase04}
and LiCuVO$_4$ which lies in a different parameter regime with
$J_1\approx -0.3\,J_2$.\cite{enderle05}
In both cases, the saturation field $h_{\mathrm{sat}}$
is within experimental reach.
A recent transfer-matrix renormalization group (TMRG)
study\cite{lu06} of the thermodynamics of 
Eq.~(\ref{eq:1.1}) was motivated by the experimental results
for Rb$_2$Cu$_2$Mo$_3$O$_{12}$.

In this paper we  study the zero-temperature phase diagram 
in a magnetic field
and the thermodynamics of Eq.~(\ref{eq:1.1}) at zero field.
The former is obtained by  a combination of a low-energy field theory and
the density matrix renormalization group (DMRG) method,\cite{white92} while  the latter
is computed by exact diagonalization (ED).
 We develop a
minimal effective field-theory description for the
region of small $J_1$ and $h < h_{\mathrm{sat}}$ and predict
the existence of an {\EO} phase. Note that at $J_1=-4 \, J_2$, the system undergoes a transition to
a FM ground state.\cite{HKNN}
The field theory predictions are verified by our DMRG results. Further, our  
ground-state phase diagram differs qualitatively
from recent mean-field predictions.\cite{dmitriev06}
In our study of thermodynamic properties,\cite{WWW} we focus
on the example of $J_1 = -3 \, J_2$ and present data
for system sizes up to $N=24$ sites.
The specific heat of LiCuVO$_4$ will be discussed elsewhere.
\cite{banks06}

%**************************************************************************
% Efective Field theory
%**************************************************************************

First we discuss an effective field theory describing the long
wavelength fluctuations of Eq.~(\ref{eq:1.1}) in the limit of strong
next-nearest neighbor interactions $J_2 \gg \abs{J_1}$.

Just below the saturation field, the problem can be mapped
onto a dilute gas of bosons.\cite{chubukov91}
This mapping, which is asymptotically exact in the subspace of two magnons,
shows that magnons bind in pairs for any $J_1 < 0$. Although the two-magnon state is not always realized as a
ground state in a magnetic field,\cite{cabra00}
Chubukov\cite{chubukov91} found that in this subspace and for
$-0.38\, J_1<J_2$, the ground-state momentum
is commensurate while for $-0.25 \, J_1<J_2<-0.38 \, J_1$, it
becomes incommensurate.
Based on the discontinuous nature of the change of momentum for the lowest
two-magnon bound state, Chubukov further predicted a first-order phase
transition between a chiral and
a dimerized  nematic-like phase.

Apart from the issue of the two-magnon states being
realized as ground states, the mapping onto a dilute gas of bosons
is controlled just near the saturation field $h \approx h_{\mathrm{sat}}$.
We apply a
complementary bosonization procedure which is controlled for
$h < h_{\mathrm{sat}}$ and confirm that the hallmark property
of the commensurate region -- pair-binding of magnons --
is universal and extends well below the saturation field.
A good starting point is the limit of
$J_2\gg \abs{J_1}$ and a finite magnetic field $h\neq 0$.\cite{cabra00}
In this limit, the system may be viewed as two
AFM chains subject to an external magnetic field and weakly
coupled by the FM zig-zag interaction $J_{1}$.  
It is well known that the low-energy effective field theory for
a single isolated spin-${1}/{2}$ chain ($J_1=0$) in a uniform
magnetic field is the Tomonaga-Luttinger liquid:\cite{LutherPeschel}
\begin{equation}
\label{SpinChainBosHam}
{\cal H} =  \frac{v}{2}\int dx \, \Big\{\frac{1}{K}(\partial_x \phi)^{2} 
+ K (\partial_x \theta)^{2}\Big\} \, .
\end{equation}
Above we have introduced a compactified scalar bosonic field $\phi$ and
its dual counterpart $\theta$, with $[\phi(x),\theta(y)] = i\Theta (y-x)$,
where $\Theta(x)$ is the Heaviside function.

The Luttinger liquid (LL) parameter $K(h)$ and spin-wave velocity $v(h)$
can be related to microscopic parameters of the lattice model $J_{2}$ and $h$
using the Bethe-ansatz solution of the Heisenberg chain in a
magnetic field.\cite{bogoliubov86,BA}
We recall here that $K(h)$ increases monotonically
with the magnetic field from $K(h=0)={1}/{2}$ to the
universal free-fermion  value $K=1$  for $h$ approaching the saturation
field $h_{\mathrm{sat}}=2\,J_2$.  
The Fermi wave vector $k_F=\frac{\pi}{2}\,(1-M)$ is determined
by the magnetization $M$. Note that we normalize the magnetization to 
$M =1$ at saturation, i.e., 
$M = 2\, \Sztot /N$ with $\Sztot = \sum_l s_l^z$.

Now we perturbatively add the interchain
coupling term to two chains, each of which  described
by an effective Hamiltonian of the form Eq.~(\ref{SpinChainBosHam})
and fields $\phi_i$, $i=1,2$. 
For convenience, we transform to the symmetric and antisymmetric combinations
$\phi_{\pm}=(\phi_1\pm \phi_2)/\sqrt{2}$ and  
$\theta_{\pm}= (\theta_1 \pm \theta_2)/\sqrt{2}$. 
In this basis and apart from terms ${\mathcal H}_{0}^{\pm}$ of the form (\ref{SpinChainBosHam}), the effective Hamiltonian 
describing low-energy properties of Eq.~(\ref{eq:1.1})
contains a single relevant coupling with the coupling
$g_1\propto J_1\ll v$:
\begin{eqnarray}
{\mathcal H}_{\rm eff} =
 {\mathcal H}_{0}^{+} +{\mathcal H}_{0}^{-} + g_1\int dx\,\cos\big(k_F+\sqrt{8\pi }\phi_{-}\big)\, ,
\label{symantisym}
\end{eqnarray}
 and the renormalized LL parameter:
\begin{equation}
\label{LL}
K_-=K(h) \, \left\{ 1 + J_1 \, K(h)/\left\lbrack \pi \, v(h)\right\rbrack \right\} \, .
\end{equation}
The Hamiltonian
(\ref{symantisym}) yields the minimal effective low-energy field theory 
describing the region $J_2\gg \abs{J_1} $ of the frustrated FM spin-${1}/{2}$
chain for $M \ne 0$.
The relevant interaction term $\cos \sqrt{8\pi}\phi_{-}$ opens
 a gap in the $\phi_-$ sector.
Since $s_{l+1}^z-s_{l}^z\sim\partial_x\phi_-$, 
relative fluctuations of the two chains are locked,
leading to pair-binding of magnons.
These bound pairs of magnons themselves are
gapless, since $s_{l+1}^z+s_{l}^z\sim\partial_x\phi_+$.
This phase was observed for an AFM  $J_1$
in Ref.~\onlinecite{okunishi03} and dubbed the `{\EO} phase'.

In addition, we confirm this picture numerically.
The simplest possible lattice model
that is described by a low-energy effective Hamiltonian
of type (\ref{symantisym})
is a spin ladder with a dominant biquadratic leg-leg interaction.\cite{nerseyan97} We compute the
magnetization curve with DMRG (results not presented here)
and verify that only even magnon sectors are realized as
ground states for all fields $h>0$.

Equation~(\ref{LL}) shows that the LL parameter $K_{-}$ decreases
with an increasing absolute value $|J_1|$ of the FM interchain coupling, in contrast to
an AFM coupling. However, the bosonization procedure
becomes inapplicable once $K_-$ vanishes. 
This signals an instability of the {\EO} phase when increasing $|J_1|$.
Moreover, we conclude that for FM
$J_1$, the {\EO} phase extends up to the saturation
field $h_{\mathrm{sat}}$ (since $K_-<1$ for $J_1 < 0$
such that there is always a relevant coupling in the
antisymmetric sector), in contrast to the AFM case.

\begin{figure}[t!]
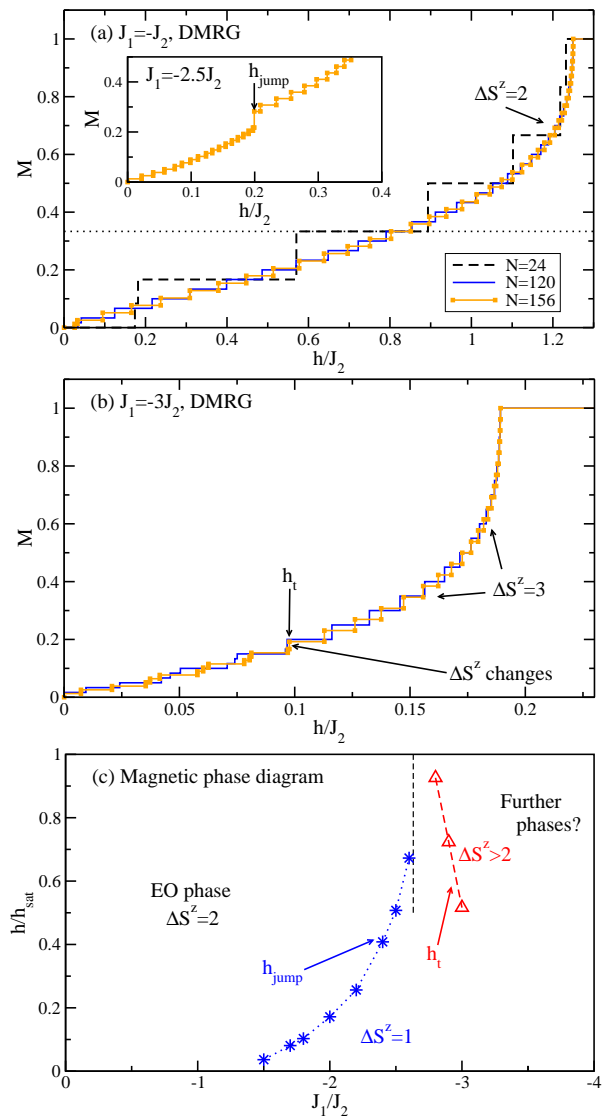

\centerline{\epsfig{file=fig1.eps,angle=0,width=0.43\textwidth}}
\centerline{\epsfig{file=fig2.eps,angle=0,width=0.43\textwidth}}
\centerline{\epsfig{file=fig3.eps,angle=0,width=0.44\textwidth}}
\caption{(Color online) 
(a), main panel (inset): Magnetization curve $M(h)$ for $J_1=-J_2$ ($J_1=-2.5\,J_2$).
The horizontal dotted line marks $M =1/3$.
(b): $M(h)$ for $J_1=-3\,J_2$. (c): Magnetic phase
diagram of the frustrated FM chain.
 The dotted line (with stars)
marks the first-order transition between the {\EO} phase and the $\Delta S^z=1$ region,
while the line $h=h_t$ (dashed, triangles) separates the $\Delta S^z=1$
region from the $\Delta S^z \ge 3$ part. Uncertainties of the transition
lines, e.g.\ due to finite-size effects, should not exceed the size of
the symbols. The fields $h_{\mathrm{jump}}$ and $h_{t}$ were extracted
from $N=156$ sites (stars) and $N=120$ sites (triangles), respectively.
The dashed, vertical line is the result of Ref.~\onlinecite{chubukov91}
($J_2\approx 0.38 J_1$).
%for the transition between the commensurate and the incommensurate region.
}
\label{fig:2d}
\end{figure}

To check this scenario and to determine the phase boundaries, we perform DMRG calculations for
up to 156 sites imposing open boundary conditions. The finite-system
algorithm\cite{white92} is used and we keep up to $350$ states.
DMRG gives direct access to the ground-state energies $E_0(\Sztot,h=0)$ 
at zero magnetic field in sub-spaces labeled by 
$\Sztot$. After shifting the ground-state energies $E_0(\Sztot,h=0)$
by a Zeeman term through
$E_0(\Sztot,h) = E_0(\Sztot,h=0) - h \, \Sztot$,
it is straightforward to construct the magnetization curve.
% We set $J_2=1$ unless stated otherwise.

We start the discussion from the limit $\abs{J_1}\lesssim J_2$.
The magnetization curves for $J_1=-J_2$ are shown in Fig.~\ref{fig:2d}~(a). 
In particular, we verify the pair-binding of magnons predicted above:
in a wide parameter range in the magnetic phase diagram ($h$ vs $J_1$),
the magnetization changes in steps of $ \Delta \Sztot =2$. This can
be observed even on  systems  as small as $N=24$,
while for an AFM interchain coupling $J_1>0$,
the formation of bound states was only reported for long
chains.\cite{okunishi03}

{}From the inset of Fig.~\ref{fig:2d}~(a), which shows data for $J_1=-2.5\,J_2$,
we conclude that a second phase emerges
at lower fields, signaled by a change of the magnetization steps from
$\Delta \Sztot =1$ to $2$ at $h=h_{\mathrm{jump}}$.
This transition  is  first order.

In contrast to the frustrated antiferromagnetic chain,\cite{okunishi03}
no indications of a $M=1/3$ plateau are found. 
We find that the width of the $1/3$ plateau as seen
on finite systems scales to zero with $1/N$.

 The magnetization curve exhibits further features when $J_1$ approaches the
transition to the FM regime, occurring at $J_1=-4\,J_2$.
The main observations from Fig.~\ref{fig:2d}~(b),
which shows $M(h)$ for $J_1=-3\,J_2$, and additional data
not displayed in the figures, are as follows. Below saturation, the steps in $M(h)$ are 
of size $\Delta \Sztot=3$ 
[see, e.g., the case of $J_1=-3\,J_2$ in Fig.~\ref{fig:2d}~(b)]. 
Upon decreasing $J_1\to -4\,J_2$, the magnetization curve becomes very
steep below saturation and the steps of $\Delta \Sztot$ may even be larger than $3$.
For instance, we find
steps of $\Delta S^z=4$ for $J_1=-3.75$ and $N=60$ below saturation.
Nevertheless, the asymptotic behavior of $M(h)$ close to the saturation field for $J_1 = -3 \, J_2$
[see Fig.~\ref{fig:2d}~(b)] is consistent with a standard square-root
singularity in $M(h)$ (see Ref.~\onlinecite{cabra00} and references therein).

Our main findings for the magnetic phase diagram 
are displayed in Fig.~\ref{fig:2d}~(c), where $h$ is
normalized by $h_{\mathrm{sat}}$.\cite{chubukov91,aligia00}
The largest part of the phase diagram belongs to the {\EO} phase, while the
transition to the region with $\Delta S^z=1$ is  first order. The position of this line, i.e., $h_{\mathrm{jump}}$,
(dotted, with stars) is consistent with results of Ref.~\onlinecite{chubukov91} in the high-field limit,
but the transition takes place at lower $h/h_{\mathrm{sat}}$ for smaller $J_1$. For larger $J_1$ and fields $h>h_{t}$, a third region
emerges, characterized by $\Delta S^z= 3$. 
Just as for AFM $J_1$,\cite{KV05}
one may speculate about chiral order in some of these regions as well as additional phases, but
substantially larger system sizes might be   needed
to fully reveal the nature of this part of the phase diagram.

%******************************************************************************************
% Thermodynamic Properties
%******************************************************************************************

Next we discuss thermodynamic properties concentrating on $h=0$.
We perform full diagonalizations to obtain all eigenvalues
and then use spectral representations to compute thermodynamic quantities,
as described in some detail for the entropy in Ref.~\onlinecite{ZhiHo}.
In order to render the Hamiltonian (\ref{eq:1.1}) translationally invariant,
we now impose periodic boundary conditions. After symmetry reduction,
the biggest matrices to be diagonalized for $N=24$ are of complex
dimension 81~752. In such high dimensions, we
use a custom shared memory parallelized Householder algorithm, while standard
library routines are used in lower dimensions.

\begin{figure}[t!]
\centerline{\epsfig{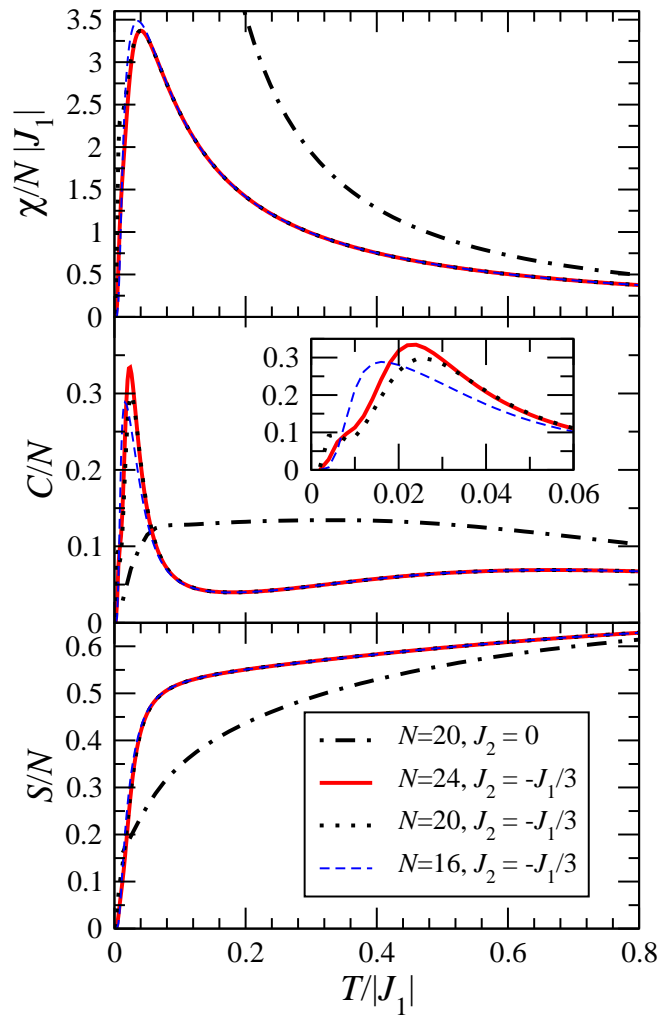}}
\caption{(Color online)
Magnetic susceptibility (top panel), specific heat (middle panel)
and entropy per site (bottom panel)
for $N=16$, $20$, and $24$ at $J_1 = -3 \, J_2$, $h=0$ in comparison to
a FM chain with $J_2=0$ and $N=20$.
Middle panel, inset: specific heat at low temperatures
 for $J_1 = -3 \, J_2$.}
\label{fig:TDm3}
\end{figure}

Figure~\ref{fig:TDm3} shows results at $J_2 = -J_1/3$ and $h=0$
for rings with $N=16$, $20$, and $24$. 
This ratio of exchange constants is 
close to values suggested for Rb$_2$Cu$_2$Mo$_3$O$_{12}$,\cite{hase04}
and the phase diagram in a magnetic field promises interesting
properties in this parameter regime.
Both the magnetic susceptibility $\chi$ and the specific heat $C$ have
a maximum at low temperatures, namely for $N=24$ at
$T \approx 0.04 \, \abs{J_1}$ in the case of $\chi$ and
$T \approx 0.023 \, \abs{J_1}$ in the case of $C$. While
these low-temperature maxima are affected by finite-size effects,
the dependence on $N$ is negligible at higher temperatures.  The specific heat exhibits a second broad maximum around
$T \approx 0.67 \, \abs{J_1}$. Such a double-peak
structure in the specific heat has already been observed for
$J_2 = -0.3\,J_1$ on a finite lattice with $N=16$ sites,\cite{TM99}
and by TMRG at $J_2 = -0.4\, J_1$.\cite{lu06}
Note that the results for $C$ of Ref.~\onlinecite{lu06}
are restricted to temperatures $T \gtrsim 0.013\, \abs{J_1}$
in this parameter regime, and the TMRG method might be plagued by convergence problems
 at low temperatures. Despite the finite-size
effects in our data at low temperatures we can clearly resolve the
low-temperature peak in $C$ (see
inset of the middle panel of Fig.~\ref{fig:TDm3}).

Our results for $\chi$ (top panel of Fig.~\ref{fig:TDm3})
differ qualitatively from those obtained for $J_2 = -0.3\, J_1$
and $N=16$ in Ref.~\onlinecite{TM99} by ED. In particular,
we  find a singlet  ground state 
for all periodic systems with $\abs{J_1} < 4\,J_2$ 
investigated, in contrast to Ref.~\onlinecite{TM99}. However, we do find good agreement with the
more recent TMRG results for $\chi$.\cite{lu06}

It is not entirely trivial to separate the low- and high-temperature features in $C$ and $\chi$ into
FM and AFM ones. Let us compare the
case $J_2 = -J_1/3$ with an unfrustrated FM chain
(Fig.~\ref{fig:TDm3} includes results for $J_2=0$, $J_1 < 0$ and $N=20$). In both cases, there is
a broad maximum in $C$ at high temperatures, although numerical values
are  different.  Concerning the low-temperature peaks
in $\chi$ and $C$,
note that, for $J_1 = -3 \,J_2$, the FM
$s=N/2$ multiplet is located at an energy of about $N\,\abs{J_1}/40$
above the $s=0$ ground state. Since this energy scale roughly agrees 
with the temperature scale of the low-temperature maxima, it is conceivable
 that they correspond to FM fluctuations
above an AFM ground state.

Finally, we note that the entropy of the frustrated FM
chain ($J_2 = -J_1/3$) is larger than that of the simple FM
chain ($J_2 = 0$) over a wide temperature range
(see bottom panel of Fig.~\ref{fig:TDm3}). Only for very low
temperatures the FM ground state leads to a bigger entropy
for $J_2 = 0$.

To summarize, 
we have studied the ground state phase diagram  of a frustrated FM chain 
in a magnetic
field and found an {\EO} phase characterized by bound pairs of magnons.
The boundary of this phase appears to be  first-order and terminates
for $h \to h_{\rm sat}$ at $J_2 \approx -0.38\, J_1$.\cite{chubukov91}
At larger FM $\abs{J_1}$, changes in the step height of the magnetization
curves signal the presence of further phases, which need to be studied in more detail. 
It would also be desirable to better understand the low-lying
excitations in the different phases and to compare to the case of the
frustrated antiferromagnetic chain.\cite{okunishi03}
Our phase diagram differs substantially from recent mean-field
predictions.\cite{dmitriev06} In particular, our DMRG data
exhibit a smooth transition to saturation for any $J_1 > -4\,J_2$,
in contrast to previous studies.\cite{dmitriev06,aligia00}
This observation may also be relevant for the transition to saturation
in the frustrated square lattice ferromagnet.\cite{shannon06}
The parameters relevant to LiCuVO$_4$\cite{enderle05}
lie well inside the {\EO} phase where the theoretical magnetization
curves are completely smooth.

Furthermore, we have discussed thermodynamic properties.\cite{WWW}
The most prominent feature for $J_2 = -J_1/3$, $h=0$ is a double-peak
structure in the specific heat.\cite{TM99,lu06}
The excitation spectrum is not reflected directly in thermodynamic
quantities, but microscopic probes such as  neutron scattering or nuclear
magnetic resonance should be able to differentiate between
gapped $\Delta S^z = 1$ excitations and gapless $\Delta S^z =2$
excitations.

%***********************************************************************************************
% Acknowledgment
%***********************************************************************************************

{\it Acknowledgments -- }
We are grateful for generous allocation of CPU time on
compute-servers at the Rechenzentrum of the TU Braunschweig
(COMPAQ ES45, IBM p575) and the HLRN Hannover (IBM p690)
as well as the technical support of J.~Sch\"ule. 
We thank M.\ Banks, D.~C.\ Cabra, S.\ Kancharla, R.\ Kremer,
R.\ Melko, and H.-J.\ Mikeska for fruitful discussions and valuable comments
on the manuscript. This work was supported in part by 
the NSF grant DMR-0443144.
%\begin{thebibliography}{99}
%*********************************************************************************************
%
\bibliography{j1ferro.bbl}

%\end{thebibliography}

%--------------------------------------------------------
\end{document}